# Anatomy of Scholarly Information Behavior Patterns in the Wake of Academic Social Media Platforms

**Hamed Alhoori**
Department of Computer Science, Northern Illinois University, DeKalb, IL, USA,
Argonne National Laboratory, Lemont, IL, USA
alhoori@niu.edu

**Mohammed Samaka**
Department of Computer Science and Engineering, Qatar University, Doha, Qatar
samaka.m@qu.edu.qa

**Richard Furuta**
Center for the Study of Digital Libraries and Department of Computer Science and Engineering, Texas A&M University, College Station, TX, USA, furuta@tamu.edu

**Edward A. Fox**
Department of Computer Science, Virginia Tech, Blacksburg, VA, USA, fox@vt.edu

**ABSTRACT**

As more scholarly content is born digital or converted to a digital format, digital libraries are becoming increasingly vital to researchers seeking to leverage scholarly big data for scientific discovery. Although scholarly products are available in abundance—especially in environments created by the advent of social networking services—little is known about international scholarly information needs, information-seeking behavior, or information use. The purpose of this paper is to address these gaps via an in-depth analysis of the information needs and information-seeking behavior of researchers, both students and faculty, at two universities, one in the U.S. and the other in Qatar. Based on this analysis, the study identifies and describes new behavior patterns on the part of researchers as they engage in the information-seeking process. The analysis reveals that the use of academic social networks has notable effects on various scholarly activities. Further, this study identifies differences between students and faculty members in regard to their use of academic social networks, and it identifies differences between researchers according to discipline. Although the researchers who participated in the present study represent a range of disciplinary and cultural backgrounds, the study reports a number of similarities in terms of the researchers' scholarly activities.

# 1. Introduction

Billions of dollars are spent each year on research and the resulting publications [1]. However, research outcomes are rarely leveraged to the fullest extent possible. This can be attributed to the fact that scholarly communities face multiple challenges. On this point, Martin M. Cummings, former director of the National Library of Medicine, summed up the situation like this: "Can a productive scientist keep abreast of a scientific literature that doubles in size every fifteen years and shows evidence of continued exponential growth during this decade? I believe that it is no longer possible to do so, even in a limited field or discipline" [2].

Establishing an understanding of researchers' scholarly activities, including the paths they take in this regard, is vital to the discovery of new strategies and techniques whereby researchers can maximize their information gains and scholarly impact. Further, a sound knowledge base pertaining to the patterns that govern these activities— herein referred to as "scholarly information behavior"—would also facilitate the efforts of libraries, publishers, and other information providers to tailor services, develop specialized collections, and build academic digital libraries and research assessment tools [3].

Over the past decade, social networking and digital library services have been widely used in academia and research environments to support researchers' scholarly activities [4][5]. Several terms are used to refer to and differentiate between these services based on the main functionalities they provide, for instance, social bookmarking for researchers [6], online or social reference management (SRM) systems [7], academic social media platforms, and academic social networks. A number of popular SRMs and academic social networks have emerged and evolved, including CiteULike [8], Zotero [9], BibSonomy [10], Mendeley [11], Academia.edu [12], and ResearchGate [13], which are used by millions of researchers worldwide.

Given that the number of scholarly products is increasing [14][15] and that numerous academic social media platforms are used during a research project's lifecycle, researchers' information needs, information-seeking behavior, and information use are not well known or understood. The purpose of the present study is to address this research gap and establish a better understanding of dynamic international scholarly information behavior by applying quantitative and qualitative methods to compare the similarities among and differences between the behavior of researchers at a university in the United States (U.S.) and at another university in Qatar. Moreover, in the present study, we investigate whether academic social networks have any effect on scholarly information behavior. By learning about the researchers' research attitudes, practices, tactics, strategies, and expectations, we will establish a basis for proposing ways to remove or overcome significant obstacles in the research process.

## 2. Related Work

Numerous studies have been conducted in a range of disciplines to understand the scholarly information behavior of various groups. The disciplinary areas explored in this regard include architecture [16], astronomy [17][18], agricultural and biological sciences [19], business [20], chemistry [21][22], computer science [23], geoscience [24], humanities [25]–[27], law [28]–[30], mathematics [31], medicine and health sciences [32]–[35], public health [36], and veterinary medicine [37]. The groups explored include the Google generation [38], undergraduate students [39][40], graduate students [41][42], scientists [43][44], engineers [45]–[47], and academic scholars [48] [49].

Information has been collected pertaining to scholarly information behavior using quantitative studies (e.g., surveys) [50]–[52], qualitative studies (e.g., interviews) [53][54], ethnographic observational studies [55][56], and combinations of these. For example, Brown [57] used a combination of email survey and content analysis methods. Further, various studies used citation analysis to study researchers' information seeking behavior and information needs [58]–[62]. Other studies investigated usability evaluation methods [63], analyzed journals and article downloads [64], and used transactional log studies [65]–[71]. Overall, diverse models have been developed to capture and analyze information-seeking behavior [72]–[74].

In an effort to better understand their information-seeking behavior, Niu et al. [75] surveyed 2,063 academic researchers in several disciplines at research universities in the U.S. The results showed clearer differences in information-seeking behavior between disciplines and between demographics than between universities. In a follow-up study, Niu and Hemminger [76] reported several factors affecting the information-seeking behavior of researchers, including demographics, psychological aspects, academic position, and discipline. Larivière, Sugimoto, and Bergeron [77] found that doctoral students cite more recently published literature than faculty members.

Scholarly use of social media [78] has been studied in blogs [79]–[82], wikis, and micro-blogging services such as Twitter [83] [84]. Recent studies have attempted to determine the influence of social media platforms on scientists and scholarly communities [85]–[90]. A few studies have investigated the effects of SRMs on scholarly communities [91]–[93]. In a study on the effects of social media tools on researchers at six universities in the United Kingdom, Tenopir, Volentine, and King [94] found that around half of the 2,000 survey respondents read, viewed, and/or participated in at least one social media platform.

A related and well-studied research area is personal information management [95]–[98], which refers to organizing and retrieving various kinds of personal collections. For example, Dumais [99] developed a system that provides a desktop personal search of information that a user has seen. Fourie [100] explored ways in which librarians engage in personal information management and reference management. In a qualitative study designed to determine the impact of electronic journals at universities in Catalan, Ollé and Borrego [101] found that the researchers tended to use either folders or bibliometric management software to organize their personal information management, or to use no identifiable information management methodology at all.

Gruzd and Goertzen [102] cited the top reasons participants gave for using social media tools related to information-gathering activities. Among these reasons were to keep up-to-date on topics [103], to follow other researchers' work, to discover new ideas or publications, to promote current research, to make new research contacts, and to collaborate with other researchers. Mandavilli [104] found that a vital reason for using social media tools is to benefit from platforms that enable discussions of scholarly output to take place in a timely manner. Jeng, He, and Jiang [105] studied a sample of users who had joined online research groups in Mendeley and found that they used the research features available more than the social features. However, most of the studies conducted with the goal of learning about scholarly information behavior are either limited to a single university campus, language, culture, or tool, or did not investigate the effects of using social media tools in academia.

## 3. Methodology

To achieve a thorough understanding of researchers' information behavior patterns, we conducted a mixed methods research study [106] whereby the qualitative aspect relied on interviews and the quantitative research relied on an online survey. The same set of questions was used for the interviews and the survey. We based our questions on categories that emerged during the literature review and observations of features of academic social networks. These questions, together with the options the participants could select as answers, are given in the Appendix.

Before the interviews and the survey were administered, seven researchers reviewed the questions to assess the efficacy and completion time required. Based on their feedback we made modifications, although these were minimal. Participation in both studies was confidential and voluntary, and the participants were informed that they were free to withdraw at any time.

To collect more information, we used semi-structured interviews conducted in the interviewees' offices. Each interview lasted between 30 and 60 minutes. We started with open-ended questions and then covered the unanswered questions. After an interviewee answered a question, s/he was given the list of other possible options, which helped us to convert the qualitative responses into quantitative data. Interviews were transcribed and coded. To analyze the interview data, we adopted a content analysis approach. The answers were analyzed and related themes were grouped by categories.

We investigated how changes in technologies available to research communities can benefit researchers, supporting their overall research progress and outcomes. In addition to collecting demographic information, we explored a number of central research questions:

- How do researchers select and use resources to search for scholarly content?
- How do researchers manage their scholarly content?
- How is collaboration taking place in scholarly communities?
- How do researchers measure the impact of research?
- Do academic social media platforms have any influence on research communities?
- What are the current information needs of researchers?
- What difficulties do researchers encounter in the research process?
- What are the similarities among and the differences between the scholarly information needs and practices of researchers at a U.S. university and those at a university in Qatar?

In the U.S., eight randomly selected faculty members from different disciplines at Texas A&M University in College Station participated in personal interviews (2 females and 6 males). Most of the interviewees supervised a research group. The interviews started with a discussion of the current practices in the research group. For the survey, invitations were sent to participants in various university departments, and the resulting samples were random and independent.

In Qatar, since the response rate for the survey was low and since few related studies have been conducted there, we focused on interviews that could provide more details. The participants were mainly faculty members from Qatar University. We randomly selected a group of 32 faculty members engaged in research, of whom 21 participated in the study (3 females and 18 males).

We refer to the first study as the *U.S. study* and to the second as the *Qatar study*. We refer to the U.S. participants as *PUX* and to Qatar participants as *PQX*, where X = {1, 2, 3,…}. In the results, we added the number of the question from the Appendix next to each finding (e.g., *QX*). We used statistical hypothesis testing techniques, principally Pearson's chi-squared test ($X^2$), analysis of variance (ANOVA), and Fisher's exact test for a small sample size.

## 4. Results

### 4.1. Survey (U.S. only)

A total of 156 researchers participated in the online survey from the U.S. study, with roles shown in Figure 1 (*Q1*). There were 124 male and 32 female respondents (*Q2*). Of these, 32 (~21%) were between 18 and 25 years old, 100 (64%) were between 26 and 34, 22 (14%) were between 35 and 54, and 2 (%1) were 55 years of age or older (*Q3*). The participating researchers represented 13 disciplines (*Q4*).

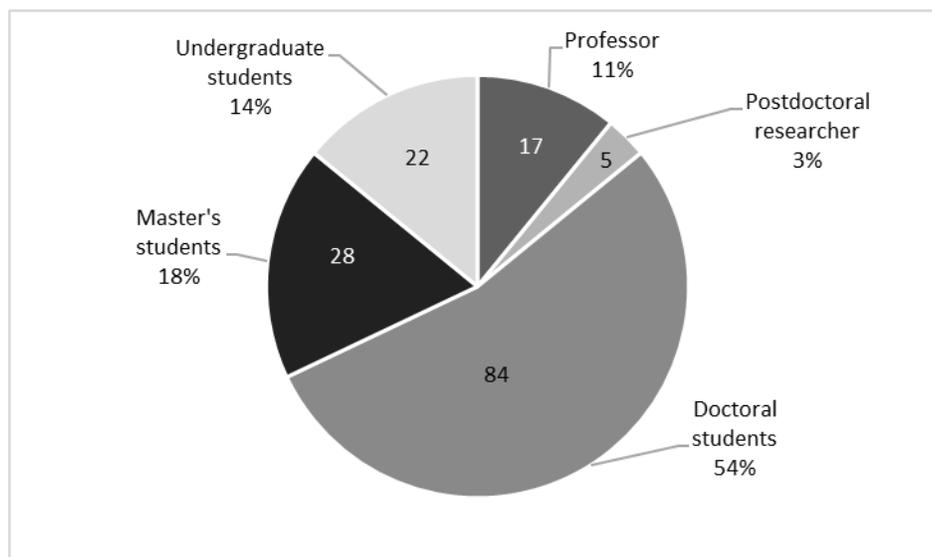

**Figure 1.** Distribution of survey participants

Surveyed participants reported that to archive the information they discover, they saved copies of articles and built personal article collections or repositories using a computer directory/folder, a reference manager, or an SRM (*Q10*). Figure 2 shows the type of personal article collection methods employed by students and faculty members (*Q1* and *Q10*). We found a significant relationship between these two factors (p < 0.001).

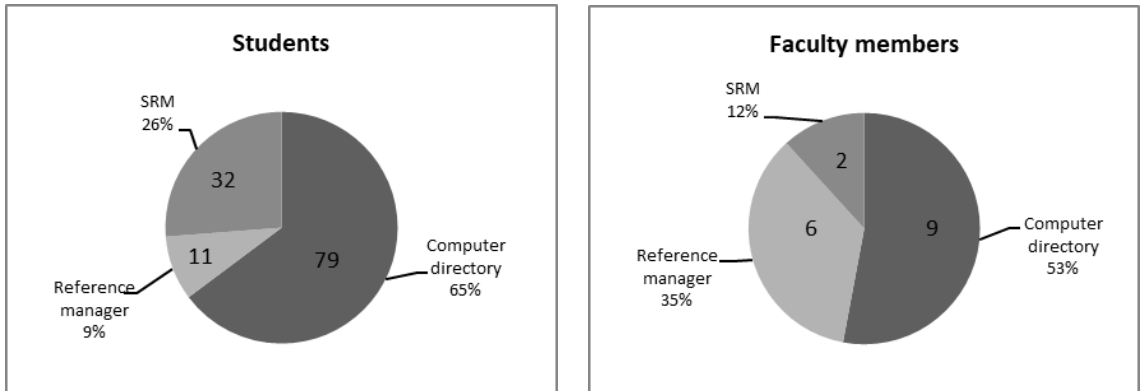

**Figure 2.** Type of personal article collection and academic status

There was no significant relationship between the type of personal article collection and gender (Figure 3, *Q2* and *Q10*).

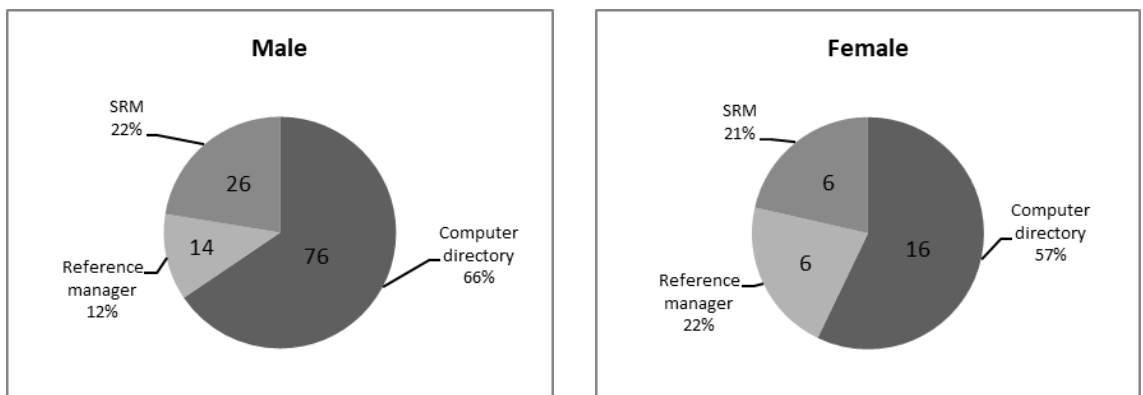

0**Figure 3.** Type of personal article collection and gender

Figure 4 shows nine disciplines and how researchers manage their scholarly article collections (Q4 and Q10). We found a significant relationship between discipline and type of personal article collection (p < 0.001). The natural science participants used SRMs as their main approach to building a personal article collection, but none of the participants in this group used a computer directory for this purpose. However, all of the economics and mathematics researchers in the study used only computer directories to build their personal article collections.

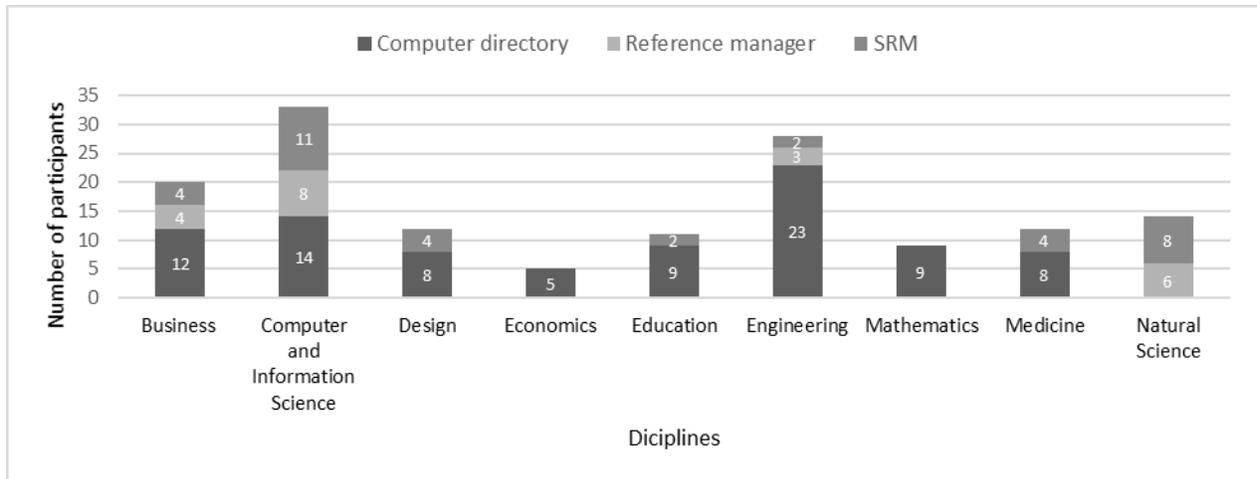

**Figure 4.** Comparison of personal article collection type usage, across nine disciplines

We considered the influence of the type of personal article collection on other scholarly activities. For example, we found that users of SRMs differ significantly from non-users of SRMs in regard to how they search for articles ($p < 0.001$) (*Q5* and *Q10*). Whereas most researchers used general or specific search engines, 40% of SRM users searched within SRMs. The participants explained that they use SRMs to search, because such platforms produce newer and more relevant results and allow them to connect with like-minded researchers (*Q6*).

The participants reported facing several difficulties in pursuing their research—i.e., a huge number of papers to filter and read, lack of knowledge in some topics, finding related work, knowing the best sequence of papers to read, and finding collaborators—as shown in Figure 5. Publication overload, which results when a researcher cannot keep abreast of the quantity of publications in his/her area of study, was a major challenge for most researchers (78%)—even for SRM users. However, there was no significant relationship between publication overload and type of personal article collection (*Q7* and *Q10*) or between publication overload and the ways in which the participants organized their articles: i.e., whether they used directories, tags, and/or visual tools [107] (*Q7* and *Q11*).

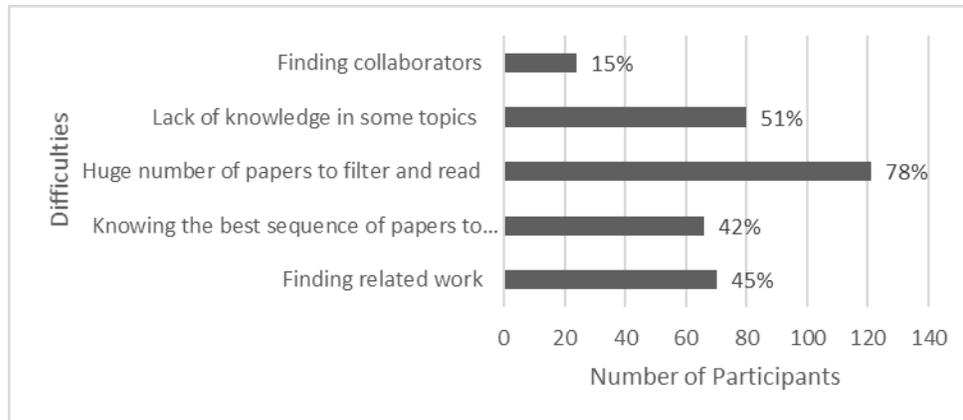

**Figure 5.** Difficulties in the research process reported by the study participants

Several participants reported that they become disoriented when navigating between articles and references, as shown in Figure 6. The results show that the participants who used directories reported becoming disoriented more often than the participants who used other approaches. We found a significant relationship between the type of personal article collection and the tendency of the survey participants to become disoriented when reading and navigating between articles ($p < 0.05$) (*Q8* and *Q10*).

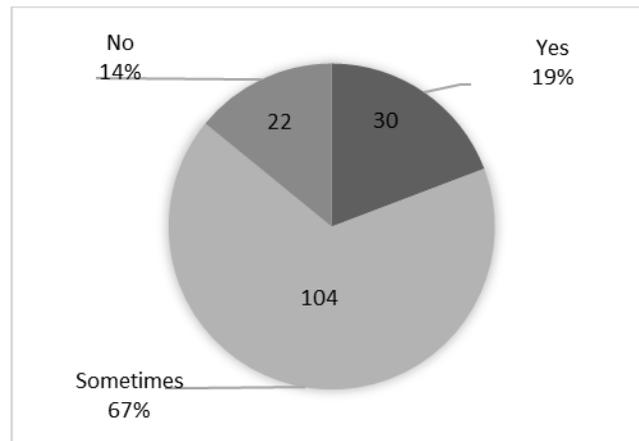

**Figure 6**. Participants' disorientation when navigating between articles

Most of the participants mentioned that they do find some related articles that would add value if cited in their completed work (Figure 7, *Q9*).

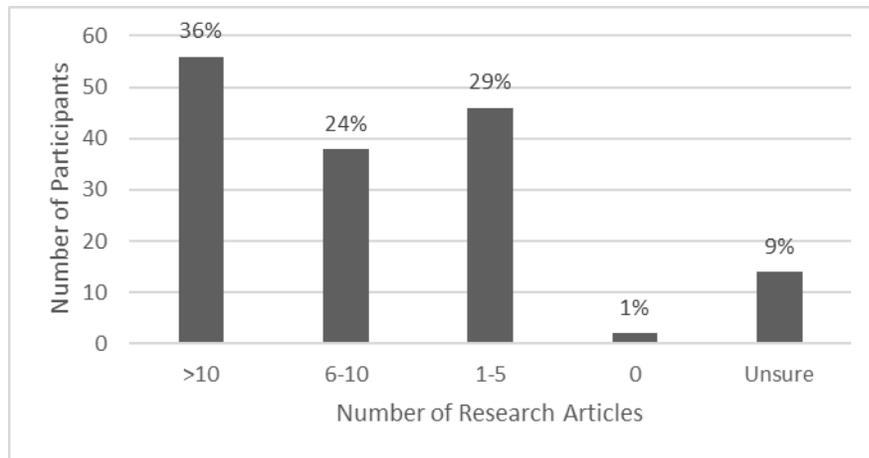

**Figure 7**. Number of related research articles that participants come across and would add value if cited in their completed work

The results show that SRM users find more articles related to their research interests than non-users of SRMs (*Q7*, *Q9*, and *Q10*). However, there was no significant relationship between using SRMs and finding related topics. Users of SRMs also used tags [108] more often than is the case for non-users of SRMs (*Q10* and *Q11*). We found a significant relationship between SRM use and tag use ($p < 0.001$). Some SRM users showed an interest in using visual tools. However, we did not find strong evidence of a relationship between using an SRM and using visual tools (*Q10* and *Q11*).

We found a significant relationship between type of personal article collection and the practice of writing notes on hard copies of articles ($p < 0.05$). The participants who wrote notes on hard copies constituted 68% of those who used directories, 50% of those who used reference managers, and only 19% of those who used SRMs (*Q10* and *Q12*). Furthermore, we found a significant relationship between the use of SRMs and the practice of making notes in an SRM ($p < 0.001$).

The first approach to retrieving articles they have read recently, for 58% of the participants, was browsing within folders, whereas 42% reported searching using keywords as their first approach (*Q13*). We found a significant relationship between the type of personal article collection and the first approach that participants used to retrieve articles they had recently read (p < 0.05). The participants who retrieved articles by searching constituted only 31% of those who used directories, 50% of those who used reference managers, and 63% of those who used SRMs (*Q10* and *Q13*). And 57.7% of the participants reported that on average they fail every week to locate at least 1-2 articles they have read previously (Figure 8, *Q14*).

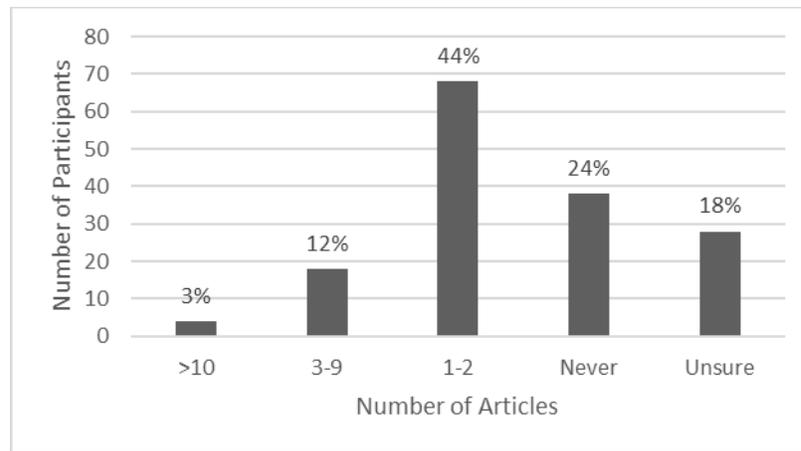

**Figure 8.** Number of research articles that participants have read previously but fail to retrieve (estimated weekly)

There was a significant relationship between the type of personal article collection and whether a researcher collaborated with other researchers (p < 0.05). Researchers who used reference managers and SRMs collaborated with more researchers than those who used directories (*Q10* and *Q15*). Many of the researchers (106; 67%) collaborated with others for one or several of the following reasons: to share and expand knowledge (55%), to make new connections (27%), to increase the possibility of securing funds (17%), to become more motivated (26%), to speed up the research process (37%), and/or to publish more (37%) (*Q17*). The researchers who did not collaborate with others (50; 32%) gave different reasons for not doing so, including being busy with their research (48%), did not see any benefits from collaborating (8%), find it hard to compile or synchronize the work (4%), not knowing other researchers with similar interests (4%), and/or other reasons (4%) (*Q16*).

The participants reported using several approaches to identify potential collaborators (*Q18*), including drawing them from among current or former research group members (77%), attending conferences (33%), noting a researcher's work being cited in several related works (23%), obtaining recommendations from an SRM (4%), and taking other researchers' suggestions (33%). The participants reported using several approaches to identify high-impact articles (*Q19*), including the number of times an article has been cited (54%), whether it has been cited in several related works (60%), the publication venue (i.e., its reputation) (61%), and recommendations from an SRM (1%).

Finally, we determined the extent to which the participants were satisfied with several scholarly activities, as shown in Figure 9. We found strong evidence that the type of personal article collection had an effect on the satisfaction of researchers when searching for articles (p < 0.001), retrieving articles (p < 0.05), and organizing articles (p < 0.05) (*Q20*).

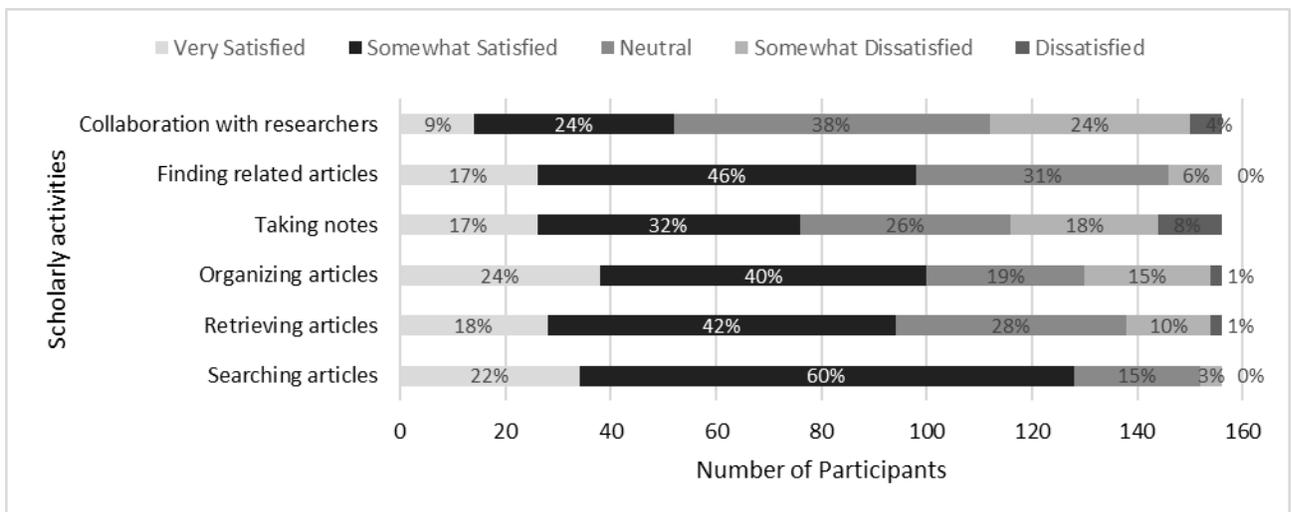

**Figure 9**. Researchers' satisfaction level with several scholarly activities

A summary of the findings is presented in Table 1 ($p < 0.05 = *$, $p < 0.001 = **$, no significance = **-**).

**Table 01.** Summary of the relationships tested

| Relationship tested in a scholarly activity | Significance |
|---|---|
| 1) SRM users and | |
|    a. searching for articles | ** |
|    b. using tags | ** |
|    c. finding related articles | - |
|    d. using visual tools | - |
|    e. making notes in an SRM | ** |
| 2) Type of personal article collection and | |
|    a. gender | - |
|    b. academic status | ** |
|    c. discipline | ** |
|    d. publication overload | - |
|    e. tendency to become disoriented | * |
|    f. practice of writing notes on hard copies of articles | * |
|    g. first approach to retrieving articles | * |
|    h. collaboration with other researchers | * |
|    i. satisfaction with searching for articles | ** |
|    j. satisfaction with retrieving articles | * |
|    k. satisfaction with organizing articles | * |

## 4.2. Interviews (U.S. and Qatar)

### 4.2.1. Searching for scholarly content

In general, the interview participants described their reliance on well-known journals, conferences, bibliographic databases, and academic digital libraries to search for articles (*Q5*).

**Noisy results.** A number of participants reported using Google Scholar, although some of these complained that this web search engine returned some articles unrelated to their search queries. Several participants complained about redundant results during the search process: *"I would like to have a way to remove the previously viewed results from my new search results or when checking for new citations. Worse than that is when I get some search results that are already stored in my articles collection or reference manager and I start to view them again since my collection is huge and I cannot remember all articles." (PU2)*

**Inefficient search.** Several participants mentioned that it was not easy to find the information they needed: *"I know the information is there, but I do not know how to reach it in a short period of time." (PU1)*. Most of the participants noted they had come across at least a few articles later that would have added value to their completed or published work had they known of the articles at an earlier point in their research (*Q9*). Others complained that sometimes they were unable to locate articles they already knew of or had even read (*Q14, Q20*): *"I usually do not succeed in finding all related work, especially those that I skim and I did not print nor read them." (PQ9)*.

**Chaining.** Following references from one article to another was shown to be a common behavior and an important discovery method for researchers in the present study: *"During my reading of an article, I jump to skim the cited articles, and around 10% of the time, I would just neglect the initial article(s) after finding more interesting and related articles to my work." (PQ4)*

**Manual search.** To keep up-to-date, some researchers noted they repeat manual searches: *"I repeat some searches from time to time and check if there are any new articles to read."* (PU5)

**Reading habits.** The participants differed in terms of their reading habits *(Q8)*, but generally agreed that they skim the paper first by reading its abstract, conclusion, or results section before deciding whether to read the entire paper. The participants generally agreed that they stop reviewing the literature when they have enough information for their purpose and/or when the content becomes repetitive.

### 4.2.2. Organizing scholarly content

In managing scholarly content *(Q10, Q11)*, researchers used single or combined approaches and some themes emerged:

**No organization.** A number of participants from both studies avoided organizing their articles, even though they regularly failed to locate articles they had read previously.

**Printing.** In organizing articles, some of the participants reported that they print the articles. When asked why they had not moved to electronic copies, they responded that they had been using this approach for a long time and did not want to jump from tool to tool: *"I print all the papers I need and organize them using authors' names. Although it may take some time to find what I need, but this way has worked for me since my graduate school."* (PU3)

**Electronic tools.** A number of participants reported being satisfied with organizing their papers and notes using computer folders and text files: *"I have been using folders to organize my papers and notes based on projects. I know all my folders, and when I need anything, I can go back to the project and to the subfolders"* (PU5). One participant even used a general organizing tool: *"I am happy using my old file organizing tool version 1.0."* (PU6)

**Reference managers.** Several participants used reference managers and shared references among their research groups. However, others, when asked why they did not use a reference manager tool, replied that they were concerned that learning to use the tool would be time-consuming and might, therefore, delay their work: *"I have used a free reference manager provided by the university library. It was good, but it needs a license and continuous updates, which delay my work, especially when I move between several places." (PU6)*. Reference managers had become an integral assessment tool for several participants. For example, one offered the following rationale for using this kind of tool: *"I have around 12,000 articles, and I am daily adding a few more. I also share some with other scholars." (PU4)*

**SRM.** Some of the participants did not know how SRMs work and refused to spend time exploring them: *"I am busy with my work and getting my tenure. I do not want to spend time using an SRM and adding friends so that I can get article recommendations" (PU3)*. A few researchers expressed regret about their lack of awareness regarding SRMs. However, some SRM users expressed concerns about the accuracy of the bibliographic data: *"I usually found some errors, missing bibliographic data or duplicate social bookmarks. So, I usually verify its data from the article's published press website" (PU8)*. Most of the researchers were aware of or had used SRMs to some extent, but one senior researcher took a position against using social networking services: *"All social media tools are distracting and produce noise, including the academic ones." (PQ16)*

**Scholarly Annotations.** Some participants reported writing notes on (*Q12*) hard copies of articles or in reference managers. Others preferred to use emails or online note-taking sites. A few even used text files and attached all saved articles, notes, or ideas to them. At least one researcher relied extensively on memory to locate a paper or a saved note: *"I have a strong memory, so I know most of my printed papers and the attached notes." (PU1)*

### 4.2.3. Research collaboration

All the faculty members collaborated on a local or an international level or both, and several were engaged in multidisciplinary collaborations (*Q15*).

**Benefits.** The faculty members usually collaborated through face-to-face meetings, communication tools (e.g., email), videoconferencing applications (e.g., Skype), and online file storage services (e.g., Dropbox): *"When conducting research in a multidisciplinary area, we are learning a new language and new skills. We try to learn what the other group is doing, and at a later point, each group will raise questions that neither group thought of before." (PQ8)*. Furthermore, the participants collaborated with other researchers to expand their knowledge and expedite their work (*Q17*). The participants reported selecting their collaborators for their expertise, reliability, and ability to work in a team (*Q18*).

**Issues.** Other participants expressed a sense of dissatisfaction with collaborating online (*Q20*): *"Even though we have regular online group meetings, we share files and results, but the collaboration is not moving as expected. Our research assistant is going to visit the other university this summer for a face-to-face collaboration." (PQ14)*

### 4.2.4. Scholarly impact

To gauge the importance of a research article (*Q19*), the participants reported reading and evaluating it. Three main themes that emerged were:

**Citation count.** Citations were considered a secondary factor in determining the value of an article.

**Publication venue**. Although the researchers sought work related to their interests in top journals, they did not consider citation-based journal rankings to be a primary measurement: *"I submitted a manuscript to a journal, and it was rejected, but I knew that the content and results were good.*

*Therefore, I resubmitted it to another journal with a higher impact factor, and it was accepted."* *(PQ14)*

**Beyond citations.** When asked how scholarly impact should be measured, one participant suggested using the PageRank algorithm: *"The impact of an article should not be measured by summing up all citations, but by knowing the reputation of the researcher who cited the article." (PU8).* Other researchers were against using citations for evaluation purposes, including a senior faculty member who pointed to the political nature of citing practices: *"The citations contain some politics in them more than science. Therefore, I think the real impact of research outcomes should be measured on how the research affected the community and human life rather than calculating a number." (PQ3)*

### 4.2.5. Research difficulties and needs

Three themes emerged in terms of research difficulties (*Q7*) and needs (*Q21*):

**Publication overload.** A number of faculty members reported suffering from publication overload. Additionally, several complained that publication overload was having a negative effect on their research assistants' work:

> *"Although I spend enough time in explaining to the research assistants the research problem, some of them get distracted by publication overload and come back with nonrelated articles." (PU7)*

> *"Some new research assistants are distracted by the huge amount of literature, and they spend a long time just to find out later that they were reading low-quality articles." (PQ10)*

After learning that several research assistants had become distracted from their assigned research task, *PQ12* found a temporary solution by creating a reading list for each student new to that role.

**Holistic solution**. The participants who used bibliographic management software sought a comprehensive solution with the ability to store all versions of articles, source codes, spreadsheets, presentations, posters, white papers, LaTeX files, Matlab files, and reports: *"I collect images of chemical formulas and store them inside documents. I also add notes near them for later retrieval." (PQ21)*. Researchers in both studies looked for advanced research tools capable of assisting them in collecting, summarizing, and analyzing the results from research articles.

**Scholarly recommendations.** In terms of receiving recommendations for articles, some of the participants wished to receive recommendations more in line with their current research direction: *"Article recommender systems usually provide recommendations related to articles that I have added to my collection a few months or years prior, while I would like to get recommendations related to my current research interests." (PQ1)*. Several researchers mentioned that they would like to receive recommendations for scholarly venues and scientific events related to their work.

Table 2 shows the resulting categories, themes, frequency values, and percentages. Figure 10 shows the connections between the main categories. A researcher is usually interested in answering a major research question, which he/she begins to address by (1) searching the literature and (2) organizing content using various tools. Next, the researcher might (3) collaborate with other researchers on conducting a study, which gives rise to scholarly outcomes. Once published, these may have a (4) scholarly impact. (5) However, there is always more work to be done: the research process and methodology are almost always limited in some way and/or the researcher(s) may encounter difficulties of one kind or another. Further research directions are, therefore, invariably needed to complement, interrogate, confirm, and/or grow from the study in order to advance the field.

Table 2: Frequency of themes in the U.S. and Qatar

| Categories | Themes | Frequency (U.S.) | Frequency (Qatar) |
|---|---|---|---|
| Searching for scholarly content | Noisy results | 3 (38%) | 6 (29%) |
| | Inefficient search | 6 (75%) | 17 (81%) |
| | Chaining | 4 (50%) | 8 (38%) |
| | Manual search | 4 (50%) | 11 (52%) |
| | Reading habits | 5 (63%) | 15 (71%) |
| Organizing scholarly content | No organization | 2 (25%) | 5 (24%) |
| | Printing | 3 (38%) | 5 (24%) |
| | Electronic tools | 5 (63%) | 14 (67%) |
| | Reference managers | 2 (25%) | 7 (33%) |
| | SRM | 3 (38%) | 4 (19%) |
| | Scholarly annotations | 7 (88%) | 16 (76%) |
| Research collaboration | Benefits | 8 (100%) | 21 (100%) |
| | Issues | 0 (0%) | 2 (10%) |
| Scholarly impact | Citation count | 6 (75%) | 18 (86%) |
| | Publication venue | 7 (88%) | 20 (95%) |
| | Beyond citations | 7 (88%) | 20 (95%) |
| Research difficulties and needs | Publication overload | 5 (63%) | 4 (19%) |
| | Holistic solution | 6 (75%) | 16 (76%) |
| | Scholarly recommendations | 3 (38%) | 7 (33%) |

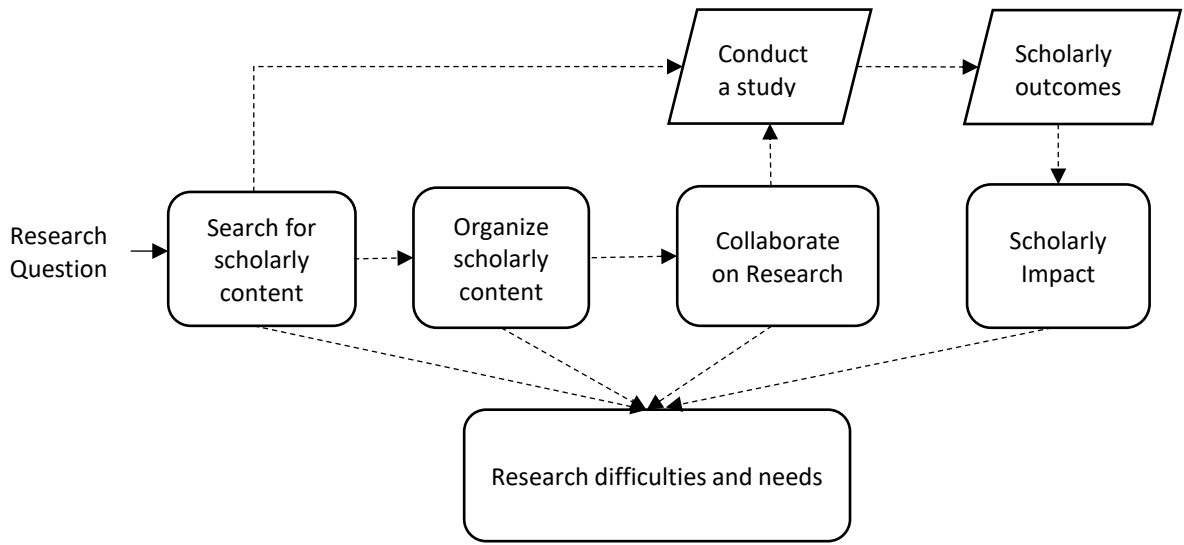

**Figure 10**: The relationship between the identified categories

## 5. Discussion

*Scholarly activities of faculty members.*

We studied the scholarly practices of 25 faculty members working in the U.S. (8 through interviews and 17 through survey) with 21 working in Qatar (all through interviews), as shown in Figure 11. When comparing these two groups, we will refer to them as *faculty interviews and survey*.

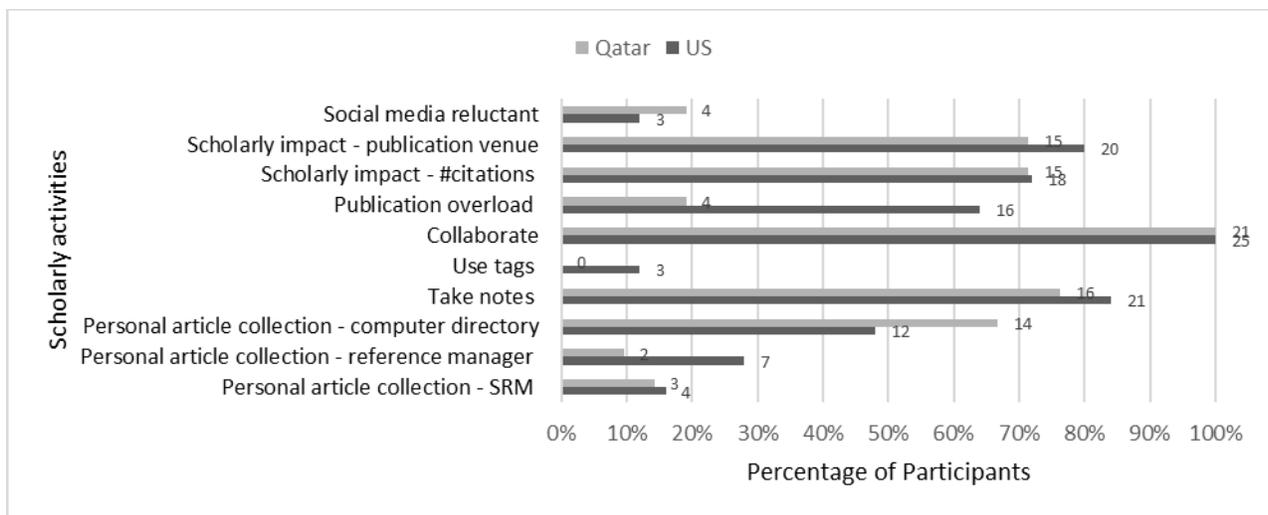

**Figure 11.** Comparison between the scholarly activities of faculty members working in the US and faculty members working in Qatar

*Finding related work.*

*How do researchers select and use resources to search for scholarly content?*

*Survey*: Although the participants used several sites or databases to search for scholarly resources, 45% reported encountering difficulties in finding related work (Figure 5, *Q7*), a result that is in line with previous findings [75], [109], [110]. Furthermore, it is unclear as to whether scholarly recommendation systems recently implemented in Google Scholar or other SRMs systems have alleviated the problem of finding related work such that further investigation in this area is needed. Whereas most researchers used general or specific search engines, 40% of SRM users searched within SRMs. The participants explained that they use SRMs to search, as such platforms produced newer and more relevant results and allowed them to connect with like-minded researchers (*Q6*). Similarly, Hallmark [111] showed that researchers in academia, government, and industry continue to develop new approaches to search for information in accordance with their needs.

*Faculty interviews and survey:* Participants from both the U.S. and Qatar used well-known scholarly venues and academic databases to search for articles. Three of the U.S. study participants used tags to organize their collections, whereas none of the participants in the Qatar study did so.

***Collaboration***.

*How is collaboration taking place in scholarly communities?*

*Survey*: Although more students than faculty members used the research assessment tools that support collaboration (e.g., SRM), not all the students engaged in collaboration during the research process, whereas all the faculty members did engage in this practice (*Q1* and *Q15*). We found that the participants who used reference managers and SRMs collaborated with more researchers than those who used directories did (*Q10* and *Q15*). Only 15% of the participants reported encountering difficulty in finding collaborators (*Q7*). However, the highest level of dissatisfaction related to scholarly activities was related to collaboration (around 28%, *Q20*).

*Faculty interviews and survey*: All faculty members from the U.S. and Qatar collaborated locally and/or internationally. Some researchers reported that they did not consider online collaboration to be as effective as face-to-face collaboration, which opens a door to investigating ways to improve the efficiency of such tools. Identifying potential collaborators usually starts from the inner circle and lab members followed by identifying other researchers at conferences, suggestions from other researchers, or reading their related work. In the present study, the reasons researchers gave for collaborating—i.e., sharing and expanding knowledge, making new connections, and having a broader impact (*Q17*)—are very similar to the reasons given in a previous study [112].

***Publication overload.***

*Survey*: The major challenge facing the participants was publication overload (78%) (*Q7*). Additionally, around 90% mentioned that they do find some related articles that would add value if cited in their completed work (Figure 7, *Q9*). Furthermore, around 57.7% mentioned that on average they do not succeed every week to locate at least 1–2 articles they have read previously (Figure 8, *Q14*).

*Faculty interviews and survey*: Publication overload affected 64% of the faculty members in the U.S. study, whereas only 19% in the Qatar study reported being affected by it. One possible explanation is that most of the participants in the Qatar study focus on selected journals and conferences, whereas those in the U.S. follow several scholarly venues and engage in multidisciplinary research. Several faculty members in the U.S. and Qatar studies noted that their research assistants were affected by publication overload. Several participants reported that they became disoriented when navigating between papers and references (*Q8*), whereas others, those who kept notes and focused on high-impact papers, did not report becoming disoriented.

***Build a personal article collection.***

*How do researchers manage their scholarly content?*

*Survey*: We found that a higher percentage of students than faculty members used SRMs to build personal article collections (Figure 3). This finding is in line with findings reported in [113], another study in which Ph.D. students were found to comprise the majority of Mendeley readers. The finding is also consistent with [77], in which it was found that Ph.D. students cite newer publications than faculty members. The finding is also consistent with results reported in a study by Emanuel [114], which showed that graduate students use Mendeley (an SRM) more than faculty members do and that faculty members use EndNote (a reference manager) more than graduate students.

*Faculty interviews and survey*: Participants in the U.S. and Qatar did not differ much in terms of using SRM or a computer directory. However, seven faculty members from the U.S. study used a reference manager, while only two in the Qatar study used them.

***Selecting research questions and shifting research interests***.

*Faculty interviews and survey*: One finding that emerged relates to selecting research questions or projects. Most researchers in the U.S. and Qatar worked on projects related to their expertise and the funding available. Other researchers reported that the quantity of work in an area influences their decision to continue working on it or shift to a new area. As one researcher stated, "*If there are many published work[s] in an area, I will move to another related area" (PQ2)*.

***Research community detection***.

In general, the participants in the interviews and survey reported that they become aware of other researchers working in similar areas by reading papers and attending conferences. However, some participants reported that they used an SRM to identify and/or to learn about researchers with similar interests.

### *Keeping up to date with research.*

*Faculty interviews and survey*: Some of the researchers reported performing repeated manual searches to find new articles, whereas others used an SRM for this purpose. The participants also used chaining to find new articles, which is consistent with the Ellis model [115] and findings reported in previous research [116], [117]. The participants generally stop reviewing the literature when they have enough information for their purpose and/or when the content becomes repetitive. This finding is in accordance with findings from studies on the information-seeking behavior of art administrators [118] and organizations [119].

### *Miscellaneous scholarly difficulties and needs.*

*What difficulties do researchers encounter in the research process? What are the current information needs of researchers?*

*Survey*: A difficulty that was faced by 51% of the participants is that they have a lack of knowledge about some topics (Figure 5). This might be related to the fact that research is becoming more interdisciplinary and requires knowledge in several areas. Despite the wide availability of multiple scholarly tools, 42% of the participants indicated that they found navigating the literature was a problem (Figure 5). Additionally, around 57.7% of the participants reported that each week on average they fail to locate at least 1–2 articles that they have read previously (Figure 8). A number of participants were not satisfied with some of their scholarly activities (Figure 9), including searching for new articles (3%), retrieving articles (11%), finding related articles (7%), taking notes (26%), and organizing articles (16%). A number of researchers mentioned that the current tools do not meet their scholarly needs.

*Faculty interviews and survey*: Several participants complained about redundant results during the search process and the need for a way to retrieve only research content that is new to them. A number of researchers reported that they were looking for better collaboration tools. Some were looking for a one-stop tool to store scholarly content, run experiments, and communicate the results. One researcher was looking for citation recommendations [120][121] based only on his current work rather than on all the research directions he has pursued to date, which has been an active area of research in the last few years.

*Social media influence.*

*Do academic social media platforms have any influence on research communities?* The results show that academic social networks have a clear impact on several research activities. For example, we found that 40% of SRM users search for articles in SRMs. Further, compared with non-users of SRMs, SRM users use more tags and are able to retrieve more articles related to their research. A greater percentage of students than faculty members used SRMs to build personal article collections.

*Social media reluctant.*

*Faculty interviews and survey*: The fact that participants from both the U.S. and Qatar were reluctant to use social media tools for scholarly purposes and to switch to new research assessment tools was consistent with results reported in other studies [122]–[124]. However, considering that only a minority are using SRMs, this reluctance is considered high. The reasons for this reluctance include learning curve, concerns about delaying research, time needed to organize and update data, accuracy of bibliographic data, insufficient benefits, and high noise and distraction level (*Q16*).

*Scholarly impact*.

*How do researchers measure the impact of research?* Citation count and publication venues remain important ways to measure scholarly impact for most interview and survey participants (*Q19*). The interview participants mentioned the need for a better approach and even proposed some alternatives, although there was no agreement on what a better approach would be.

*Awareness and misconceptions*.

*Faculty interviews and survey:* Some researchers were not aware of academic social media platforms or some of their features. There were also some misconceptions about SRMs: In particular, three participants from the U.S. and two from Qatar expected to devote extensive time to learning how to use an SRM, whereas the principles can be learned in minutes, and the two participants from the U.S. and two from Qatar thought it was necessary to add friends to get recommendations; however, this is not the case. A number of studies have shown that not only are some researchers unaware of or unfamiliar with some of the resources, services, and electronic search tools available to them through libraries but also that they generally do not discuss their information needs with librarians [57], [125]–[127]. To raise researchers' awareness of the tools available to them, workshops and online tutorials [128][129] have been provided to support researchers' activities, on topics such as the use of specific tools [130] (e.g., bibliographic management software).

*Disciplinary differences*.

*Survey*: we found a relationship between disciplines and type of personal article collection (Figure 4). This result is in accordance with Niu et al.'s [75] finding of differences in information-seeking behavior among disciplines.

*Faculty interviews and survey*: Insufficient data were available to check for disciplinary differences.

*Similarities and differences*.

*What are the similarities among and the differences between the scholarly information needs and practices of researchers at a U.S. university and those at a university in Qatar?*

*Faculty interviews and survey:* We found some differences between the U.S. and Qatar studies, and similarly some previous studies also note differences in researchers' scholarly activities [42], [75], [76]. Liao, Finn, and Lu [131] found differences between international and American graduate students in information-seeking behavior using an online survey at Virginia Tech of 315 respondents. They investigated how students begin their search for information using several choices which included: classmates, professors, or librarians; reference resources; textbooks or lecture notes; Addison library catalog; library e-resources (electronic journals, databases, and electronic theses and dissertations); Internet; or other. They found that around 50% of international students begin their research by searching on the Internet and that e-resources is their second choice (16.5%), whereas 40.6% of U.S students conduct their first search using library e-resources, while the Internet is their second choice (29%). We also found similarities between participants in both of our studies. Our findings show that some participants used similar scholarly resources, collaborated with other researchers, and used more than one method to build personal article collections and write notes.

## 6. Conclusion and Future Work

Although large-scale scholarly digital libraries provide more enhanced services, tools, and methods to researchers, little is known about the ways in which researchers explore the research landscape, discover relevant information, and satisfy their information needs. In this paper, we investigated current practices and scholarly activities on an international level in the social media age. We compared the scholarly information behavior and information needs of researchers at a university in the U.S. and of researchers at a university in Qatar. By revealing several significant relationships, the survey we conducted deepens our overall understanding of scholarly attitudes. We found a number of similarities in regard to the behavior and needs of researchers in both studies. We also found that SRMs are important to researchers in their efforts to find and organize scholarly articles and to connect with other researchers.

These studies showed that publication overload continues to affect researchers. The researchers who had built a personal article collection were more satisfied with their information needs than those who had not built a collection of this nature. Overall, we found that researchers are not fully utilizing scholarly information sources and tools. Moreover, even with all the advances in scholarly and social platforms, researchers reported that their information needs are not being fully met.

Current academic digital libraries and SRMs are based on a one size fits all approach, such that newer implementations should be designed to address the specific needs of researchers in a range of disciplines. Many researchers become comfortable with the tools they are using, hence new technologies must come with very clear benefits if researchers are to become motivated to try them.

As a next step in extending the research presented herein, we plan a quantitative study with a larger sample of researchers that will include an investigation of the specific needs of researchers according to discipline. George et al. [109] found that nearly all graduate students (96%) reported that academics influence their research and information seeking. We will investigate whether SRMs have any significant effect on research groups in building online collaborative research communities, serendipity, and temporal information searching behavior [132]. Collaborative information seeking and social information seeking [133] have been studied and modeled with the goal of furthering our understanding of group work and group activities. We intend to investigate the effects of SRMs on the research process and to develop a collaborative research model of dynamic strategies using supervised and unsupervised machine-learning techniques [134][135]. We will investigate scholarly information behavior among researchers producing or dealing with non-English content. Additionally, we plan to investigate how social media can build and affect a research culture using various recommendations and text analytics approaches [138][139].

**Acknowledgements**

This publication was made possible by NPRP grant # 4–029–1–007 from the Qatar National Research Fund (a member of Qatar Foundation). The statements made herein are solely the responsibility of the authors. This work was supported in part by the Office of Advanced Scientific Computing Research, Office of Science, U.S. Department of Energy, under Contract DE-AC02–06CH11357. An earlier version of the initial work was presented at a TPDL conference [136] and an ICADL conference [137].

**Appendix**

Dear researcher,

We are conducting a research study with an objective to better understand the dynamic scholarly activities of researchers. Your participation is voluntary and your response is highly encouraged, valued, and any information that you provide will be confidential and anonymous.

Many thanks for your time and assistance.

*Questions:*

Q1 What is your current position?

- Professor
- Postdoctoral
- Doctoral student
- Master's student
- Undergraduate student
- Other _____________

Q2 What is your gender?

- Male
- Female
- Other

Q3 What is your age range?

- Under 18
- 18–25
- 26–34
- 35–54
- 55–64
- 65 or over

Q4 What is your academic discipline?

- ○ Arts and Humanities
- ○ Business
- ○ Computer and Information Science
- ○ Design
- ○ Economics
- ○ Education
- ○ Engineering
- ○ Law
- ○ Mathematics
- ○ Medicine
- ○ Natural Science
- ○ Social Sciences
- ○ Other

Q5 What sites or databases do you use to search for scholarly resources?

- ❑ Google Scholar
- ❑ University library
- ❑ Within reference manager software (e.g., EndNote)
- ❑ Social reference managers (e.g., Mendeley, CiteULike, or Zotero)
- ❑ Specialized sites or databases (e.g., IEEE/PUBMED)
- ❑ Others _________________

**Answer** If What sites or databases do you use to search for scholarly resources? Social reference managers (e.g., Mendeley, CiteULike, Zotero) *Is Selected*

Q6 What made you use a social reference manager while searching?

- ❑ Accurate bibliographic data
- ❑ Newer articles
- ❑ More relevant results
- ❑ Knowing users with similar interests
- ❑ Others _______________________

Q7 What difficulties do you encounter in the research process?

- ❑ Finding related work
- ❑ Knowing the best sequence of papers to read
- ❑ Huge amount of papers to filter and read
- ❑ Lack of knowledge in some topics
- ❑ Finding collaborators

❏ Others ____________

Q8 Do you get distracted while reading and moving between articles and references?

○ Yes
○ Sometimes
○ No
○ Unsure

Q9 How many related articles did you come across that would add value if cited in your completed work (estimated an average for a single work)?

○ More than 10
○ 5–10
○ 1–5
○ 0
○ Unsure

Q10 What is your main method of storing research articles that you have read?

○ Computer folders/directory
○ Reference manager (e.g., EndNote)
○ Social reference managers (e.g., Mendeley, CiteULike, Zotero)
○ None
○ Others___________________

Q11 How do you organize the research articles that you have read?

❏ Folders
❏ Tags
❏ Visual tools
❏ Don't organize
❏ Others ___________________

Q12 Where do you keep notes about articles you read?

- ❏ On printed paper
- ❏ Text file
- ❏ Reference manager (e.g., EndNote)
- ❏ Specialized sites (e.g., ACM binder)
- ❏ Social reference managers (e.g., Mendeley, CiteULike, Zotero) (5)
- ❏ Emails
- ❏ Others ____________________
- ❏ Don't take notes

Q13 What is your first approach to retrieving articles that you read recently?

- ❍ Browse within folders
- ❍ Search using keywords

Q14 How often do you not succeed in retrieving articles that you have read previously (estimate weekly)?

- ❍ >10 articles
- ❍ 3–9
- ❍ 1–2
- ❍ Never
- ❍ Unsure

Q15 Have you collaborated with other researchers?

- ❍ Yes (1)
- ❍ No (2)

*Answer* If Have you collaborated with other researchers? No *Is Selected*

Q16 Why don't you collaborate with other researchers?

- ❏ Don't see any benefits from collaborating
- ❏ Busy with my research
- ❏ Don't know them
- ❏ Don't trust them
- ❏ Others ____________________
- ❏ Waste of time
- ❏ Hard to compile/synchronize the work

*Answer* If Have you collaborated with other researchers? Yes *Is Selected*

Q17 Why do you collaborate with other researchers?

- ❏ Publish more
- ❏ Increase possibility of getting funds
- ❏ Motivate each other
- ❏ Share and expand knowledge
- ❏ Make new connections
- ❏ Others ____________________
- ❏ Speed up the work

Q18 How do/would you know potential collaborators?

- ❏ Current or former research group members
- ❏ Attending conferences
- ❏ Researchers' work cited in several related works
- ❏ Recommendations from a social reference manager
- ❏ Other researchers' suggestions
- ❏ Others ____________________

Q19 How do you identify the high-impact articles in a research area?

|  | High impact articles |
|---|---|
| # Citations of articles | ❏ |
| Cited in several related works | ❏ |
| Publication venues | ❏ |
| Recommendations from social reference manager | ❏ |
| Don't know | ❏ |
| Others | ❏ |

Q20 How would you rate your approaches for:

|  | Very Satisfied | Somewhat Satisfied | Neutral | Somewhat Dissatisfied | Dissatisfied |
|---|---|---|---|---|---|
| 1. Searching articles | ❍ | ❍ | ❍ | ❍ | ❍ |
| 2. Retrieving articles | ❍ | ❍ | ❍ | ❍ | ❍ |
| 3. Organizing articles | ❍ | ❍ | ❍ | ❍ | ❍ |
| 4. Taking notes | ❍ | ❍ | ❍ | ❍ | ❍ |
| 5. Finding related articles | ❍ | ❍ | ❍ | ❍ | ❍ |
| 6. Collaboration with researchers | ❍ | ❍ | ❍ | ❍ | ❍ |

Q21 Do you have any other specific needs in your work that you would like to be included in research tools?